\documentclass[final,epj]{svjour}

\usepackage{epsfig}
\usepackage{graphicx}

\begin{document}

\title{On the Rapid Increase of Intermittency in the Near-Dissipation Range of Fully Developed Turbulence}
\subtitle{}

\author{Laurent Chevillard \inst{1 \and 2} \and  Bernard Castaing \inst{1} \and Emmanuel L\'ev\^eque \inst{1}}
\institute{Laboratoire de physique, \textsc{Cnrs}, \'Ecole normale sup\'erieure de Lyon, France \and Laboratoire
\'Ecoulements G\'eophysiques et Industriels, \textsc{Cnrs}, Universit\'e Grenoble 1, France}

\mail{Laurent.Chevillard@ens-lyon.fr}
\titlerunning{Intermittency in the near-dissipation range of turbulence}
\authorrunning{L. Chevillard et al.}

\abstract{ Intermittency, measured as $\log
\left({F(r)}/{3}\right)$, where $F(r)$ is the flatness of velocity
increments at scale $r$, is found  to rapidly increase as viscous
effects intensify, and eventually saturate at very small scales.
This feature defines a finite intermediate range of
scales between the inertial and dissipation ranges, that we shall
call near-dissipation range. It is argued that intermittency is
multiplied by a universal factor, independent of the Reynolds
number $Re$, throughout the near-dissipation range. The
(logarithmic) extension of the near-dissipation range varies as
$\sqrt{\log Re}$. As a consequence, scaling properties of velocity
increments in the near-dissipation range strongly depend on the Reynolds number.
\keywords{fully developed turbulence -- intermittency --
small-scale dissipative effects} }

\PACS{ 05.45.-a  47.27.-i  47.27.Eq  47.27.Gs 47.27.Jv}

\maketitle

\section{Introduction}

Statistics of developed turbulence are commonly investigated by
means of (longitudinal) velocity increments $\delta v(r)$ across a
distance, or scale, $r$. At $r \approx L_0$, where $L_0$
represents the characteristic scale of the stirring forces
(the integral scale of turbulence), fluid motions are statistically
independent and the probability density function (pdf) of $\delta
v(L_0)$ is found nearly Gaussian. At smaller scales, intrinsic
non-linear fluid dynamics operate and turbulent motions become
intermittent; fluid activity comes in intense locally-organized
motions embedded in a sea of relatively quiescent and disordered
eddies (see \cite{meneguzzi,orszag} for first numerical
indications). As a consequence, the pdf of $\delta v(r)$ develops
long tails and becomes strongly non-Gaussian. Deviations from the
Gaussian shape may be quantified by the flatness, defined as
$$ F(r) \equiv \frac{\langle \delta v(r)^4 \rangle}{\langle \delta
v(r)^2 \rangle^2}\mbox{ .}$$ For a centered Gaussian distribution
$F=3$; as long tails develop $F$ increases. $F(r)/3$ may therefore
be roughly thought of as the ratio of intense to quiescent fluid
motions at scale $r$. In that sense, we shall assume in the
following that $\log \left({F(r)}/{3} \right)$ provides a
quantitative measure of intermittency \cite{morf}.

\begin{figure}[htbp]
\epsfig{file=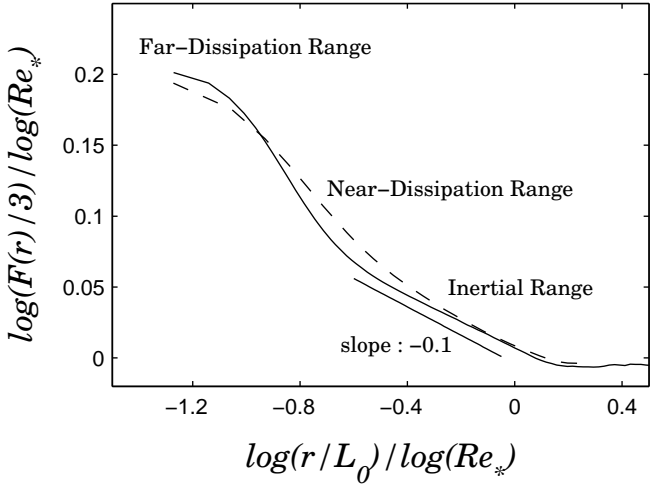,width=\hsize} \caption{\label{flatness_fig} Scale dependence of the flatness of
longitudinal velocity increments for two different turbulent flows: (solid line) a turbulent jet ($R_\lambda =
380$) \cite{baudet} and (dashed line) a direct numerical simulation ($R_\lambda=140$) \cite{numeric}. The control
parameter $Re_*$ is defined as $Re / R_*$, where $Re$ denotes the (usual) Reynolds number; the empirical constant
$R_* \approx 56$  \cite{gagne,lohse}.
 Appendix \ref{an:Rstar} is devoted to the (modified) Reynolds
number $Re_*$.
%The Reynolds number
%$Re$ is given by $R_\lambda^2/15$, where $R_\lambda$ denotes
%the Taylor micro-scale based Reynolds number \cite{frisch}.
}
\end{figure}

The normalized (to the Gaussian value) flatness is plotted as a
function of the scale ratio ${r}/{L_0}$ for two turbulent flows in
Fig. \ref{flatness_fig}. Experimentally, a particular attention
has been paid to the size of the hot-wire probe and to the
signal-to-noise ratio; small-scale velocity fluctuations are
expected to be suitably resolved \cite{helium}.  The
``instantaneous Taylor hypothesis" {(see \cite{PinTaylor}
for details)} has been used to estimate spatial velocity
increments and reduce modulation effects \cite{morf,helium}.
Details about the (standard) numerical integration of the Navier-Stokes
equations can be found in \cite{numeric}. We observe at
scales $r \geq L_0$, $F(r) \simeq 3$, in agreement with the picture of disordered
fluid motions: There is no intermittency, since the
flatness $F(r)$ is  independent of the scale $r$ and (almost) equal to the Gaussian value $F=3$.
At smaller scales, $F(r)$
displays a power-law dependence on $r$: Intermittency grows up
linearly with $\log(1/r)$. This scaling behavior is inherent to the
\emph{inertial} (non-linear) fluid dynamics and refers to the so-called
inertial range. The exponent $\zeta_F \simeq -0.1$ is found very
consistent with already reported values for homogeneous and
isotropic turbulence \cite{scalings}. Interestingly, $F(r)$
exhibits a rapid increase as viscous effects intensify, and
eventually saturates at very small scales. This rapid increase of
intermittency, which occurs over a range of scales that we shall
call the near-dissipation range, is the main concern of this
article. We shall here argue how intermittency in the
near-dissipation range is related to the build-up of intermittency
in the inertial range; the Reynolds-number dependence of this
phenomenon will be also addressed.

There has been a considerable amount of works on intermittency in
the inertial range (see \cite{frisch} for a review).
Dissipation-range intermittency has received much less attention.
In 1967, Kraichnan conjectured ``unlimited intermittency" for the
modulus of velocity Fourier modes at very high wavenumbers
\cite{kraichnan}. Although no proof was explicitly established for the
Navier-Stokes equations, Frisch and Morf provided in 1981 strong
mathematical arguments (occurrence of complex-time singularities)  \cite{morf}
in support of Kraichnan's conjecture. Following
a previous study carried out by Paladin and Vulpiani in 1987
\cite{paladin}, Frisch and Vergassola suggested in 1991 that
multifractal (local) exponents $h$ of velocity increments, $\delta v(r)
\sim r^h$, are successively turned-off as viscous effects
intensify \cite{vergassola}: As the scale $r$ decreases,
only the strongest fluctuations (low $h$) survive while the others
are extinguished by the viscosity. This mechanism reinforces the
contrast between intense and quiescent motions, and thus provides
a phenomenological explanation for the increase of intermittency
in the near-dissipation range. However, as the remaining intense
motions concentrate on a smaller and smaller fraction of the
volume, this approach again predicts ``unlimited intermittency"
for the velocity increments $\delta v(r)$, in the limit of vanishing scale $r$.

As mentioned above, our experimental and numerical data indicate
that intermittency, measured by the flatness of velocity
increments, does exhibit a blow up in the beginning of the
dissipation range but eventually saturates in the far-dissipation
range. At this point, it should be mentioned that a spurious
limitation of intermittency may stem from a lack of accuracy (or
resolution) in velocity measurements or numerical simulations.
However, a special care has been taken here to reduce this effect
\cite{helium}. In the following, we will argue that the observed
saturation of intermittency is not an artefact, but refers to some
peculiar properties of turbulence at very small, dissipative
scales.

\section{A multiplicative cascade description of intermittency}

In the present study, the issue of intermittency in the
dissipation range is reconsidered. The saturation of the flatness
in the limit of vanishing scale $r$ is recovered by assuming that
the velocity field is smooth (regular) in quiescent-flow regions
(as already suggested in \cite{nelkin,meneveau}): $\delta v(r)$ is
not zero but behaves as $r$ in these regions;
 $\delta v(x,r) \approx  r\partial_x v(x)$. This key
assumption is here recast in a multiplicative approach of
velocity-increment statistics along scales, as brought forward by
Castaing {\it et al.} in \cite{CasGag90}. We shall then
demonstrate that it is possible to gain quantitative results,
without ad hoc parameters, on dissipative-range intermittency: The
amplification of intermittency in the near-dissipation range and
the extension of the near-dissipation range are explicitly
estimated as a function of the Reynolds number.

The build-up of intermittency along the whole range of excited
scales is related to the distortion of the pdf of $\delta v(r)$.
In order to account for this distortion, let us formally introduce
a random independent\emph{ multiplier} $\beta({r}/{L_0})$, connecting the
statistics of $\delta v(r)$ at scales $r$ and $L_0$:
\begin{equation}
\delta v(r) = \beta(\frac{r}{L_0}) \times \delta v(L_0) \quad
\textrm{for } r \leq L_0~. \label{Eqbeta}
\end{equation}
The integral scale $L_0$ is taken as the reference scale. Eq.~(\ref{Eqbeta}) should be understood in the
statistical sense, i.e.,  the pdf of $\delta v(r)$ equals the pdf of $\beta({r}/{L_0}) \times \delta v(L_0)$ (see
\cite{AmbBro99} for a Markovian description). The multiplier $\beta({r}/{L_0})$ is considered as a positive random
variable. This approach therefore restricts to $\vert \delta v(r) \vert$ or to the symmetric part of the pdf of
$\delta v(r)$; the skewness effects are beyond the scope of the present description.

From Eq. (\ref{Eqbeta}), it can be established that
$$
P_r( \delta v ) = \int G_{r,L_0}(\log \beta) P_{L_0} ( \frac{
\delta v}{\beta}) \frac{\textrm d \log \beta}{\beta}~,
$$
where $P_r$ and $G_{r,L_0}$  denote respectively the pdf of
$\delta v(r)$ and of $\log \beta({r}/{L_0})$. The pdf of $\delta
v(L_0)$ may be considered as Gaussian, as mentioned in the
introduction. Once $P_{L_0}$  is known, $P_r$ is  fully determined
by $G_{r,L_0}$.

The so-called propagator kernel $G_{r,L_0}$
\cite{helium,CasGag90,arneodo,delour} is characterized by the
whole set of coefficients $C_n({r}/{L_0})$,  defined by the
expansion
\begin{equation}
\langle (\beta(\frac{r}{L_0}))^p \rangle=\exp\left(
\sum_{n=1}^\infty C_n(\frac{r}{L_0}) \frac{p^n}{n!} \right)
\quad\textrm{for all }p. \label{cum}
\end{equation}
By construction, $C_n({r}/{L_0})$ is the n$^\textrm{th}$-order
cumulant of the random variable $\log \beta(r/L_0)$. For our
purpose, we shall only focus on the first two cumulants: The mean
\begin{equation}
C_1(\frac{r}{L_0}) \equiv \langle \log  \beta(\frac{r}{L_0})
\rangle \label{Eqc1}
\end{equation}
and the variance
\begin{equation}
C_2(\frac{r}{L_0}) \equiv \langle \log^2 \beta(\frac{r}{L_0})
\rangle - \langle \log \beta(\frac{r}{L_0}) \rangle^2~.
\label{Eqc2}
\end{equation}

Experimentally, higher-order cumulants of $\log \beta({r}/{L_0})$ are found very small compared to
$C_1({r}/{L_0})$ and $C_2({r}/{L_0})$ \cite{arneodo,delour}. This motivates our main interest in the mean and
variance of $\log \beta({r}/{L_0})$ \cite{PhDChev}. However, the exact shape of the propagator kernel $G_{r,L_0}$
is not relevant for the following analysis: Our arguments apply to the mean and the variance but does not require
$G_{r,L_0}$ being Gaussian. This point should be unambiguous. We shall demonstrate that considering the mean and
the variance of $G_{r,L_0}$ is valuable in order to describe the amplification of intermittency in the
near-dissipation range.

What can be said about $C_1({r}/{L_0})$ and
$C_2({r}/{L_0})$?
First of all, it is straightforward to get
from Eq. (\ref{Eqbeta}) and Eq. (\ref{cum}):
$$
\langle |\delta v(r)|^p \rangle = K_p\sigma ^p \exp\left(
\sum_{n=1}^\infty C_n(\frac{r}{L_0}) \frac{p^n}{n!} \right)
$$
by assuming that $\delta v(L_0)$ is a zero-mean gaussian variable of variance $\sigma^2$ (i.e., $\langle |\delta
v(L_0)|^p\rangle = K_p\sigma ^p$).

-- In the inertial range,
\begin{equation}\label{eq:structfonc}\langle |\delta v(r)|^p
\rangle = K_p\sigma^p\left(\frac{r}{L_0}\right)^{\zeta_p}\mbox{ .}\end{equation} This is the \emph{postulate} of
universal power-law scalings \cite{frisch,K41}. It follows that $C_1({r}/{L_0})$ and $C_2({r}/{L_0})$ behave as
linear functions of $\log ({r}/{L_0})$:
\begin{eqnarray}
C_1(\frac{r}{L_0}) &=& c_1 \log( \frac{r}{L_0}) \quad \textrm{and} \nonumber\\
 C_2(\frac{r}{L_0}) &=& c_2 \log (\frac{r}{L_0}) \quad \textrm{in the inertial
 range,}\label{eq:defL0}
\end{eqnarray}
where $c_1$ and $c_2$ are universal constants \cite{arneodo}. The
departure from the Kolmogorov's linear scaling law $\zeta_p =
{p}/{3}$ is directly related to $c_2<0$: $$\zeta_p = c_1 p + c_2
\frac{p^2}{2} +\cdots$$ In our framework, the build-up of
intermittency (along the inertial range) is related to the
increasing width of $G_{r,L_0}$ with the decreasing scale $r$,
stating that the second-order cumulant $C_2(r/L_0)$ increases as
$r$ decreases.

-- In the far-dissipative range, velocity increments are proportional to the scale separation $r$, which leads to
\begin{eqnarray}
C_1(\frac{r}{L_0}) &=& \log (\frac{r}{L_0}) + C_1^\mathrm{diss.} \quad \textrm{and} \nonumber\\
C_2(\frac{r}{L_0}) &=& C_2^\mathrm{diss.} \quad \textrm{in the far-dissipation range.} \nonumber
\end{eqnarray}
The constants $C_1^\mathrm{diss.}$ and $C_2^\mathrm{diss.}$ a priori depend on the Reynolds number, here defined
as
\begin{equation}
R_e = \frac{\sigma  L_0}{\nu},
\end{equation}
where $L_0$ is the integral scale pointed out by Eq. (\ref{eq:defL0}), $\sigma$ denotes the standard deviation of
$\delta v({L_0})$ and $\nu$ is the kinematic molecular viscosity.

-- Finally, the inertial-range and far-dissipation-range behaviors
of $C_1({r}/{L_0})$ and $C_2({r}/{L_0})$ match in the
near-dissipation range. We will see that this matching is (very)
peculiar.

\begin{figure}
\begin{center}
\epsfig{file=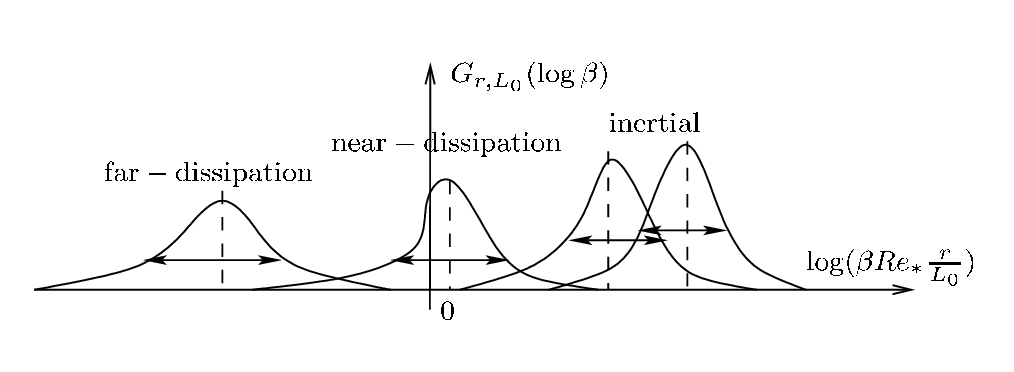,width=\hsize}
\caption{\label{schema0}Sketch of the distortion of the pdf of
$\log \beta({r}/{L_0})$  along scales $r$,  respectively
in the inertial, near-dissipation and far-dissipation ranges. The
width of the pdf increases rapidly when crossing the
near-dissipation range from right to left; with
decreasing scale $r$.}
\end{center}
\end{figure}

\section{The near-dissipation range}

Following  Paladin and Vulpiani \cite{paladin}, one considers that viscous effects at a given scale $r$ only
affect the fluctuations of $\delta v(r)$, for which the\emph{ local Reynolds number} is smaller than a certain
constant $R_*$. In the classical phenomenology of turbulence $R_*$ is fixed to unity \cite{frisch}, but for our
purpose,  $R_*$ is kept  as an empirical constant.

In our multiplicative cascade framework, the previous hypothesis writes
$$ \frac{\sigma\beta(\frac{r}{L_0})  r}{\nu}\le R_* \mbox{ .}$$
This condition is equivalent to
\begin{equation}
\beta(\frac{r}{L_0})  Re_*\left( \frac{r}{L_0} \right) \le 1
\mbox{ ,} \label{diss_effect}
\end{equation}
where $Re_*$ denotes the (modified) Reynolds number
\begin{equation}
Re_* \equiv \frac{R_e}{ R_*}\mbox{ .} \label{ree}
\end{equation}
The subscript $*$ indicates that $ R_*$ is not a priori equal to
unity; this point is clarified in the Appendix \ref{an:Rstar}.

\begin{figure}
\begin{center}
\epsfig{file=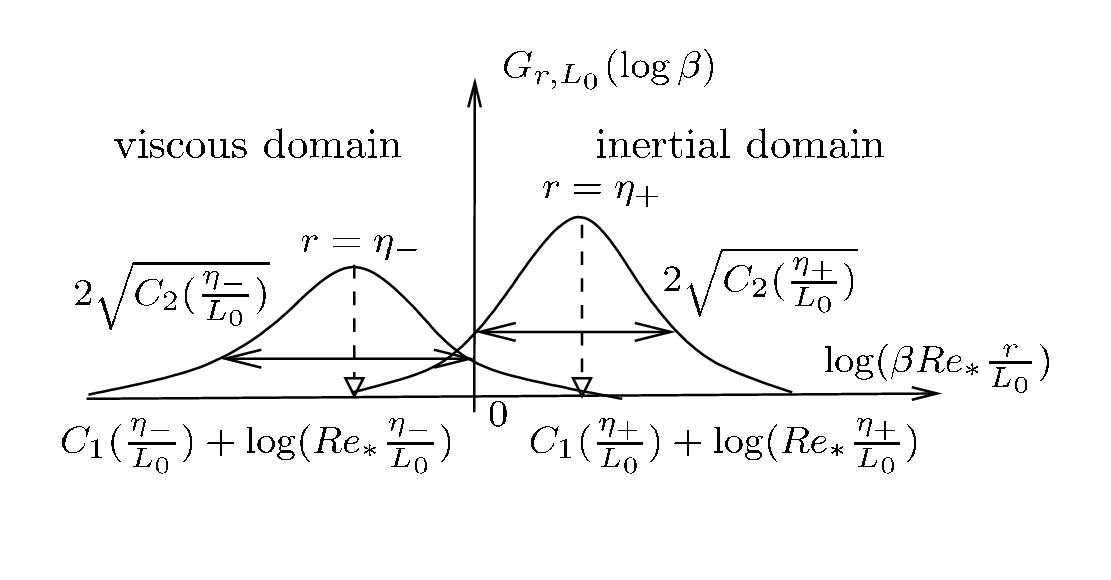,width=\hsize} \caption{ The
near-dissipation range is given by  $\eta_- < r < \eta_+$. The
characteristic scales $\eta^+$ and $\eta^-$ are defined by
 Eqs. (\ref{Eqndr1}) and (\ref{Eqndr2}): The scale $\eta^+$
may be viewed as the smallest limiting scale for which the
propagator $G_{r,L_0}$  is not affected by viscous effects; only
non-linear dynamics prevail at $r>\eta^+$. The scale $\eta^-$
may be viewed as the largest limiting scale for which the
propagator $G_{r,L_0}$ does not undergo non-linear effects; only
viscous dynamics prevail at $r<\eta^-$.}  \label{schema}
\end{center}
\end{figure}

The propagator kernel $G_{r,L_0}$ is sketched in Fig.
\ref{schema0} for various scales $r$ (the integral scale $L_0$ is
fixed) as a function of $\log(\beta Re_* {r}/{L_0})$: $G_{r,L_0}$
moves from right to left as the scale $r$ decreases. At a given
scale $r$,  fluctuations $\beta({r}/{L_0})$ which satisfy Eq.
(\ref{diss_effect}) undergo dissipative effects. As viscosity
strongly depletes these affected fluctuations, a significant
stretching of the \emph{left} tail of $G_{r,L_0}$ is expected (see
Fig. \ref{schema0}). Viscous effects then lead to a strong
distortion of $G_{r,L_0}$;  a significant increase of $C_2(r/L_0)$
follows. This argument provides a qualitative explanation of the
increase of intermittency due to (non-uniform) viscous effects.
Within this picture, the near-dissipation range may be viewed as
the range of scales $r$ marked by the entering of $G_{r,L_0}$ in
the viscous domain, and the leaving of $G_{r,L_0}$ from the
inertial domain (see also Fig. \ref{schema}).

In order to pursue a more quantitative analysis, an explicit
definition of the near-dissipation range is required. To do so,
let us introduce the two characteristic scales $\eta_+$ and
$\eta_-$ (see Fig. \ref{schema}) given respectively by
\begin{eqnarray}
C_1(\frac{\eta_+}{L_0}) + \log (Re_*\frac{\eta_+}{L_0}) &=&
\sqrt{C_2(\frac{\eta_+}{L_0})}
\label{Eqndr1}\\
\textrm{and}\quad  C_1(\frac{\eta_-}{L_0}) + \log
(Re_*\frac{\eta_-}{L_0}) &=& - \sqrt{C_2(\frac{\eta_-}{L_0})}~.
\label{Eqndr2}
\end{eqnarray}
According to our previous considerations, $\eta_+$ may be seen as
the scale marking the entering of $G_{r,L_0}$ in the viscous
domain, and $\eta_-$ as the scale marking the leaving of
$G_{r,L_0}$ from the inertial domain. In other words, $\eta_- <r <
\eta_+$ may be considered as the near-dissipation range; this will
be our (explicit) definition.

It is natural to match the inertial-range and dissipation-range
behaviors of $C_1({r}/{L_0})$ at the characteristic scale $\eta$,
for which the propagator kernel, approximately centered around its
mean value, extends equally over the inertial and viscous domains
(see Figs. \ref{schema0} and \ref{schema}). This statement writes
$$
\left< \log \left( \beta(\frac{\eta}{L_0})  Re_*\left(
\frac{\eta}{L_0} \right) \right)\right> = 0 \mbox{ ,}
$$
and yields
$$C_1(\frac{\eta}{L_0}) + \log(Re_*\frac{\eta}{L_0}) = 0\mbox{ .}$$
By assuming that the intermittency correction on $c_1$ is very
small, so that $c_1 \approx {1}/{3}$ according to Kolmogorov's
theory \cite{K41}, one obtains that $\eta$ coincides with the
notorious Kolmogorov's scale $\eta_K$, based on the modified
Reynolds number $Re_*$:
\begin{equation} \eta  = \eta_K = L_0(Re_*)^{-3/4}
\label{kd}\mbox{ .}\end{equation}
Furthermore, $C_1^\mathrm{diss.}= {1}/{2} \log Re_*$. From { Eqs.}
(\ref{Eqndr1}), (\ref{Eqndr2}) and (\ref{kd}), it follows
$$
\frac{\sqrt{C_2(\frac{\eta_+}{L_0})}}{\sqrt{C_2(\frac{\eta_-}{L_0})}}
= \frac{\frac{4}{3}\log (\frac{\eta_+}{L_0}) + \log Re_* }{-2\log
(\frac{\eta_-}{L_0})-\frac{3}{2}\log Re_*} =
\frac{\frac{4}{3}\log(\frac{\eta_+}{\eta})}{2\log(\frac{\eta}{\eta_-})}
\mbox{ ,}
$$
and by considering that $G_{\eta_-,L_0}$ results from the
distortion of $G_{\eta_+,L_0}$, one obtains {(see Appendix
\ref{eq:annGeom} for a \emph{kinematic} proof)}
\begin{equation}\label{eq:C2etaplus}
2\sqrt{C_2(\frac{\eta_+}{L_0})} \approx \frac{4}{3} \log
(\frac{\eta_+}{\eta_-}) \mbox{ }\end{equation} and
\begin{equation}\label{eq:C2etamoins}
2\sqrt{C_2(\frac{\eta_-}{L_0})} \approx 2\log
(\frac{\eta_+}{\eta_-})\mbox{ .}
\end{equation}
These results finally yield $ \log ({\eta_+}/{\eta})  =  \log
({\eta}/{\eta_-})$ and $$\eta = \sqrt{\eta_+ \eta_-}~.$$ In
logarithmic coordinates the dissipative scale $\eta$, given by Eq.
(\ref{kd}), lies at the center of the near-dissipation range:
$\eta$ separates the inertial range and the far-dissipation range,
as originally proposed by Kolmogorov. This is a first result
concerning the near-dissipation range.
% C'EST UN PREMIER RESULTAT, ON LE MET EN VALEUR

\begin{figure}[htb]
\begin{center}
\epsfig{file=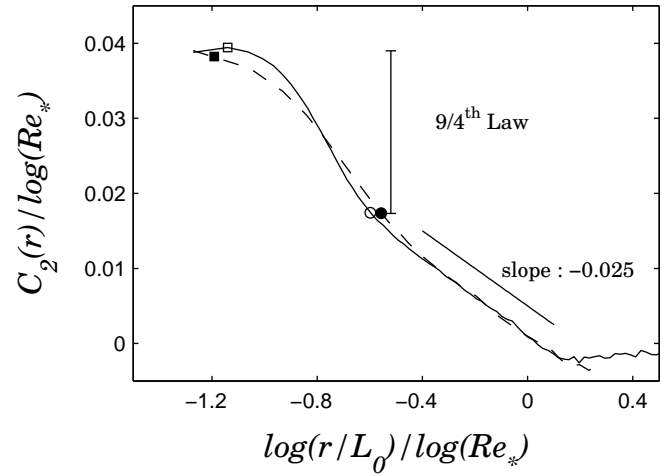,width=\hsize}
\end{center}
\caption{\label{c2}Scaling behavior of $C_2({r}/{L_0})$ for the
turbulent jet and the numerical simulation. In the inertial range,
$C_2(r) = c_2 \log(r/L_0)$ with $c_2 \approx -0.025$. The two
dissipative scales $\eta_+$ and $\eta_-$, defined by Eqs.
(\ref{Eqndr1}) and (\ref{Eqndr2}), are indicated for both flows.
The ``amplification law" between $C_2({\eta_-}/{L_0})$ and
$C_2({\eta_+}/{L_0})$ is very well satisfied. We  observe also
that the Kolmogorov's dissipative scale $\eta$, given by Eq.
(\ref{kd}), lies approximatively at the center of the
near-dissipation range: $\eta_- < r < \eta_+$. }
\end{figure}

\section{The amplification law}

From the above computation, one can derive the ``amplification law"
\begin{equation}
{C_2(\frac{\eta_-}{L_0})} = \frac{9}{4}~
{{C_2(\frac{\eta_+}{L_0})}}~, \label{law}
\end{equation}
which characterizes the increase of intermittency in the near-dissipation range. The ${9}/{4}$ factor relies on
the (reasonable) approximation that $c_1 \approx {1}/{3}$. This does not mean at all that intermittency is ignored
in our approach; it is just assumed here that the value of the parameter $c_1$, entering in the description, can
be considered very close to its Kolmogorov value. Anyhow, the same reasoning could be pursued by keeping $c_1$ as
a free parameter. In that case, the multiplicative factor would express as $4/(1+c_1)^2$. Experimentally, one
finds $c_1 = 0.37 \pm 0.01$ \cite{CasGag90}.

Interestingly, the amplification of intermittency in the near-dissipation range is universal, independent of the
\linebreak (very high) Reynolds number. Let us also remark  that this ``amplification law'' may serve as a useful
benchmark for the (experimental or numerical) resolution of the finest velocity fluctuations; one should be able
to differentiate  (true) viscous damping and spurious filtering (see \cite{kadanoff} for such a debate).

\begin{figure}[h]
\begin{center}
\epsfig{file=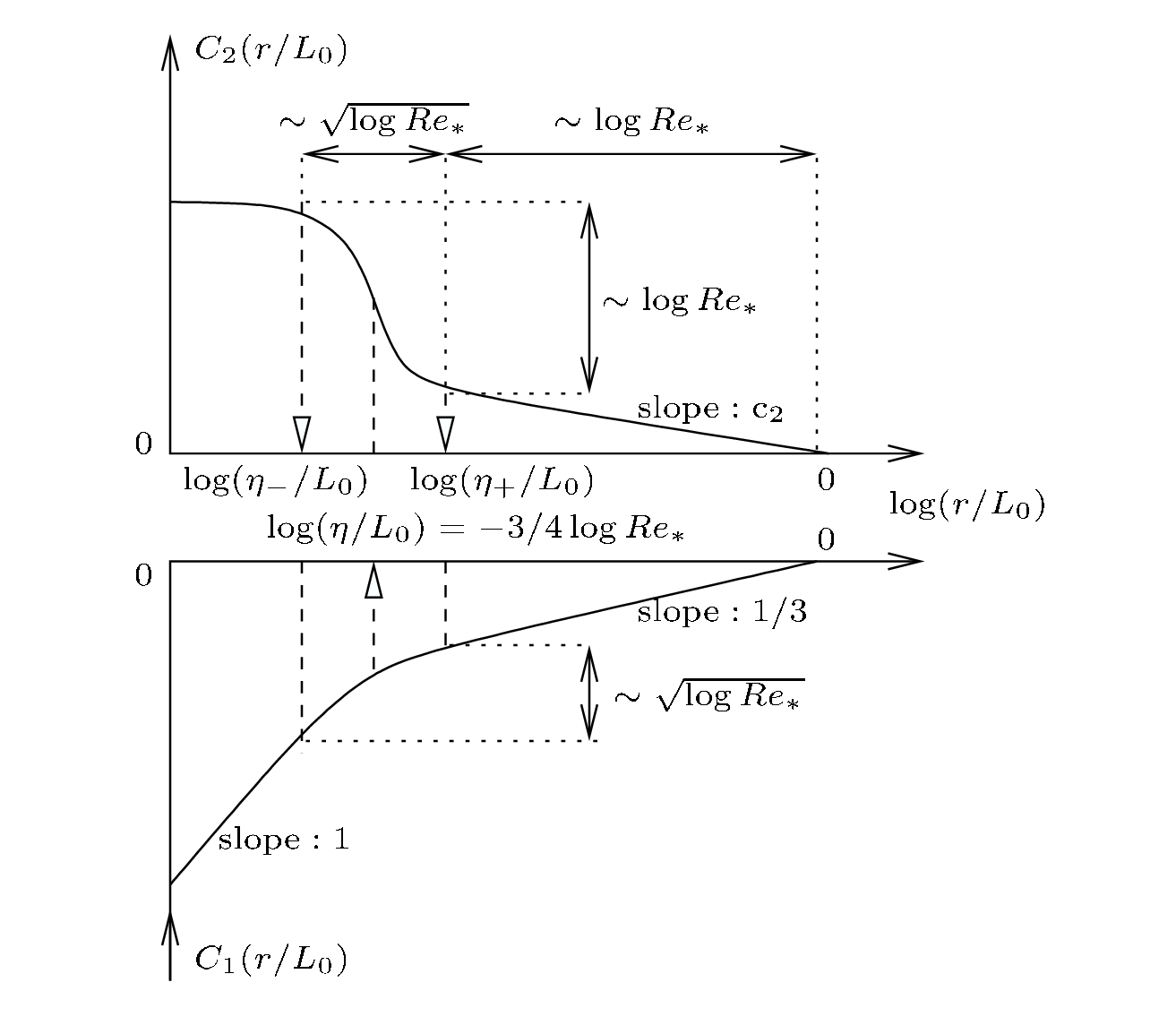,width=\hsize}
\caption{\label{schema_ess}Sketch of the scaling behavior of
$C_1({r}/{L_0})$ and $C_2({r}/{L_0})$ at high Reynolds
number.}
\end{center}
\end{figure}

\section{A unified picture of intermittency}

At very high Reynolds number, $\sqrt{C_2({r}/{L_0})}$ becomes
negligible compared to $\log Re_*$ in the near-dissipation range. It follows from
Eqs.~(\ref{Eqndr1}), (\ref{Eqndr2}) and the ``amplification
law'' that  $\eta_+ \approx \eta_- \approx \eta$ at
leading order in $\log Re_*$. This yields
\begin{eqnarray}
C_2(\frac{\eta_-}{L_0})-C_2(\frac{\eta_+}{L_0}) &\propto&  \log Re_* \nonumber\\
\textrm{and}\quad C_1(\frac{\eta_-}{L_0})-C_1(\frac{\eta_+}{L_0}) &\approx&  -\frac{2}{3} \log (\frac{\eta_+}{\eta_-} )\nonumber\\
\mathrm{with} \quad \log(\frac{\eta_+}{\eta_-} )&\propto&   \sqrt{
\log Re_*}~. \label{eq:extendneardiss}
\end{eqnarray}

These behaviors are sketched in Fig. \ref{schema_ess}. One obtains that the (logarithmic) extension of the
near-dissipation range ($\sim \sqrt{\log Re_*}$) becomes negligible compared to the extension of the inertial
range ($\sim \log Re_*$). This is consistent with the tendency observed in Figs. \ref{flatness_fig} and \ref{c2}.
At this point, it should be emphasized that $\eta_+$ can not be assimilated to the Taylor's microscale $\lambda$.
Indeed, $\log (\eta_+/\eta) \propto \sqrt{\log Re_*}$ while on the contrary $\log(\lambda/\eta) \propto \log
Re_*$.

Velocity increments follow a
log-infinitely divisible law \cite{saito92,Nov94} in the inertial
range, since all cumulants are proportional to a same function
of the scale --- $ C_p({r}/{L_0})=c_p \log ({r}/{L_0})$
for all  $p$ ---  but this log-infinitely divisibility can not pertain in the
near-dissipation range, according to the sketch in Fig. \ref{schema_ess}.

\begin{figure}[h]
\begin{center}
\epsfig{file=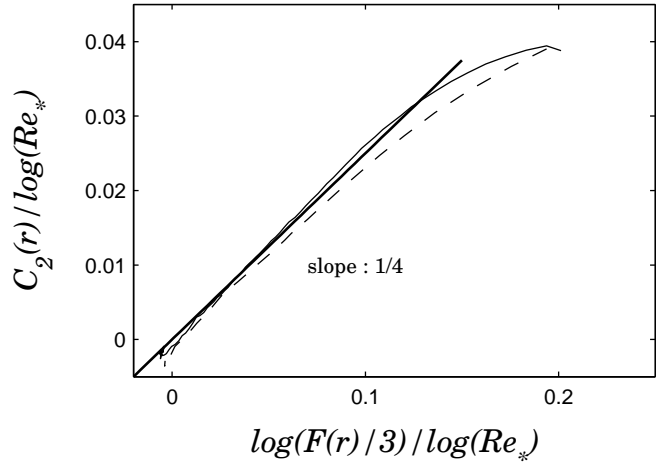,width=\hsize} \caption{We observe that Eq. (\ref{f3c2}) is well satisfied in the
inertial range (for the turbulent jet and the numerical simulation). A small departure is observed at small
scales, when viscous effects intensify. This departure may be attributed to the growing of cumulants of order
$p\geq 3$, neglected in Eq. (\ref{f3c2}).
 }\label{f3c2_fig}
\end{center}
\end{figure}

Intermittency has been related to $\log({F(r)}/{3})$ in the
beginning. Using Eq. (\ref{Eqbeta}), one can derive the exact
equation
\begin{equation}
\log \left(\frac{F(r)}{3} \right) =  \sum_{p \geq 2}
C_p(\frac{r}{L_0}) \left(\frac{2^{2p}}{p!} - \frac{2^{p+1}}{p!}
\right), \label{flat}
\end{equation}
where $C_p({r}/{L_0})$ is the $p$-th order cumulant of $\log
\beta({r}/{L_0})$. At leading order, this yields
\begin{equation}
\log\left (\frac{F(r)}{3} \right) =   4 C_2(\frac{r}{L_0}) +\cdots,
\label{f3c2}
\end{equation}
which suggests that $\log ({F(r)}/{3} )$ and $C(r/L_0)$  should
behave in a very comparable way. Fig. \ref{flatness_fig} and Fig.
\ref{c2} are indeed very similar.
 One may thus consider that
$C_2({r}/{L_0})$ provides an alternative measure of intermittency, less intuitive than $\log \left({F(r)}/{3}
\right)$ but physically more tractable (as demonstrated by this study). A specific test of Eq. (\ref{f3c2}) is
provided in Fig. \ref{f3c2_fig}.

Before concluding this study, let us mention that it is quite direct to generalize the previous analysis to
$N$-order velocity increments: $$\delta^{(1)} v(r) = v(x+r)-v(x),$$ $$\delta^{(2)} v(r) \equiv
v(x+r)-2v(x+\frac{r}{2})+v(x), \quad \textrm{etc.}$$ Inertial-range scalings $\langle |\delta^{(N)}v(r)|^p \rangle
\sim r^{\zeta_p}$ are preserved  (as argued in \cite{arneodo}) but $\langle |\delta^{(N)}v(r)|^p \rangle \sim
r^{Np}$ in the far-dissipative range. As a result, the definition of the near-dissipation is unchanged but the
``amplification law" becomes \begin{equation} {C_2^{(N)}(\frac{\eta_-}{L_0})} =
\frac{9}{4}\left(\frac{(N+1)}{2}\right)^2 {{C_2^{(N)}(\frac{\eta_+}{L_0})}}. \label{lawN} \end{equation} The
amplification factor depends on the order $N$ of the velocity increment.  This feature allows us to
discriminate the inertial range and the near-dissipation range: Inertial-range scalings do not depend on $N$,
while on the opposite, near-dissipation-range scalings depend drastically on $N$. Finally, the amplification
factor in Eq. (\ref{lawN}) diverges with $N$, tending toward Kraichnan's view of unlimited intermittency (for the
fluctuations
 of velocity Fourier modes) as $N \to \infty$ \cite{kraichnan}.

\section{Conclusion}

A unified picture of velocity-increment intermittency, from the integral scale to the smallest (excited) scales of
motion, is proposed. It is explicitly stated how far-dissipation range and inertial-range intermittencies match in
the near-dissipation range.  Especially, a universal ``amplification law'' determines how intermittency of
velocity gradients is linked to the build-up of intermittency in the inertial range. The results are found in good
agreement with our experimental and numerical observations.

Beyond these precise results, this study indicates that there are some peculiar and interesting physics around the
Kolmogorov's dissipative scale of turbulence. Such issue may be of great importance, for instance, in the
modelling of mixing properties of turbulence, which mainly rely on the behavior of gradient fields
\cite{chev_prl}.

Finally, we would like to insist on the fact that this description leads to predictive results which could be used
as tests for the suitable resolution of (very) small-scale fluctuations, to distinguish ``probe effects'' and true
viscous damping. Relations between this study and the so-called property of \emph{Extended Self-Similarity}
\cite{ess} should deserve some interests as well.

\begin{acknowledgement}
We thank C. Baudet, A. Naert, B. Chabaud and coworkers for
providing us the experimental data. We are grateful to J.-F.
Pinton and A. Arneodo for critical comments. Numerical simulations
were performed on  a IBM SP3 supercomputer at the CINES,
Montpellier (France).
\end{acknowledgement}

\appendix

\section{The (modified) Reynolds number $Re_*$}\label{an:Rstar}
{The empirical constant $R_*$, which is abusively fixed to unity in the classical phenomenology of turbulence
\cite{frisch}, may be linked to the Kolmogorov's constant $c_K$ (see \cite{yeungzhou} and references therein).

In Kolmogorov's 1941 theory, $c_K$  can be defined through the second-order velocity structure function:
$$
\left< (\delta v(r))^2\right > = c_K \left < \epsilon \right>
^{2/3} r^{2/3} \mbox{ ,}
$$
where $\left < \epsilon \right>$ denotes the mean dissipation rate. Here, intermittency corrections are obviously
omitted. By the use of  Eq. (\ref{eq:structfonc}), one can write
$$
\langle (\delta v(r))^2\rangle = \sigma^2 \left(
\frac{r}{L_0}\right)^{2/3}\mbox{ ,}
$$
which yields
\begin{equation}\label{eq:cksigma}
c_K = \frac{\sigma^2}{\langle \epsilon \rangle ^{2/3}
L_0^{2/3}}\mbox{ .}
\end{equation}
In this \emph{monofractal} description, the near-dissipation range is degenerate and reduces to the Kolmogorov's
scale $\eta_K$. The second-order moment of velocity gradient expresses as
\begin{equation}\label{eq:calcdiff}
\left < (\partial_xv)^2\right > = \frac{\left < (\delta
v(\eta_K))^2\right >}{(\eta_K)^2} \mbox{ .}
\end{equation}
By assuming homogeneous and isotropic turbulence, the mean
dissipation rate writes $\langle \epsilon \rangle =15\nu \langle
(\partial_xv)^2\rangle$. By combining the Eqs. (\ref{eq:cksigma})
and (\ref{eq:calcdiff}), together with the definition of $\eta_K$
given by Eq. (\ref{kd}), one gets
\begin{equation}
c_K = \left( \frac{R_*}{15}\right)^{2/3} \quad \textrm{or} \quad
R_* = 15 c_K^{3/2}\mbox{ .} \label{ck}
\end{equation}
Eq. (\ref{ck}) indicates that $R_*$ is eventually much greater than unity. Following \cite{gagne}, the empirical
value $R_*\approx 56$ corresponds to $c_K \approx 2.4$. This value is in good agreement with experimental and
numerical estimations $c_K \approx 2$ \cite{yeungzhou}. }

\section{Kinematic proof of Eqs. (\ref{eq:C2etaplus}) and (\ref{eq:C2etamoins})}
\label{eq:annGeom}
\begin{figure}
\begin{center}
\epsfig{file=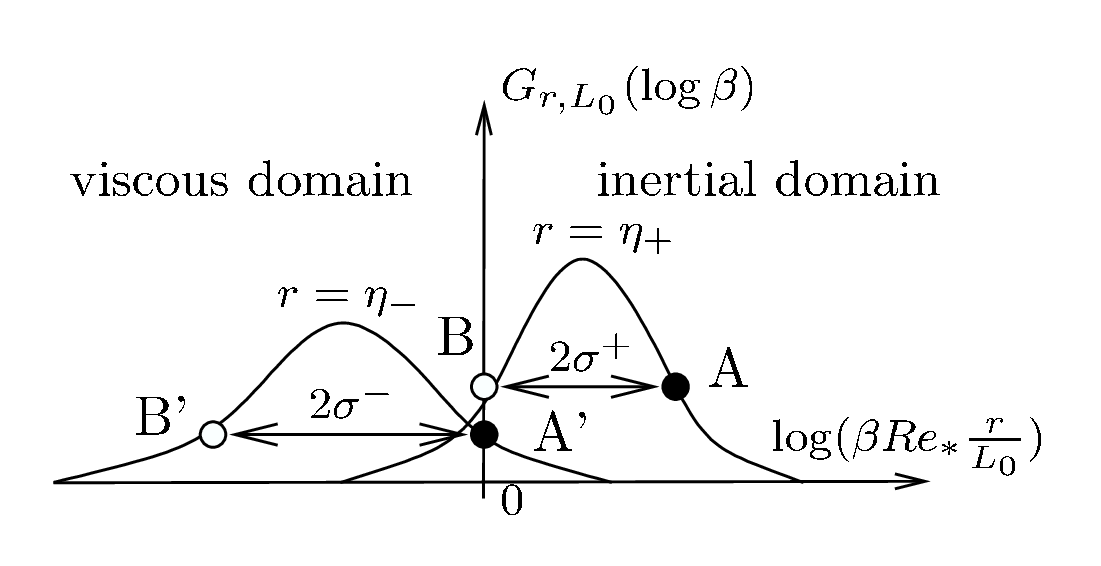,width=\hsize} \caption{ {When the scale $r$ decreases, the propagator kernel
$G_{r,L_0}$  moves from right to left ($\sigma$ denotes the standard deviation). At scale $r=\eta_+$, the points
$A$ and $B$ are defined by the standard deviation around the mean. As $r$ decreases from $\eta_+$ to $\eta_-$, the
points $A$ and $B$ move respectively to $A'$ (within the inertial domain) and to $B'$ (within the viscous
domain).} } \label{schemabis}
\end{center}
\end{figure}

{We  provide a kinematic proof of Eqs. (\ref{eq:C2etaplus}) and (\ref{eq:C2etamoins}) with the
help of Fig. \ref{schemabis}:

 \noindent
--- on the one hand, one derives from Eq. (\ref{Eqndr1}) that the distance
$$x_{A}-x_{A'}=2\sqrt{C_2(\frac{\eta_+}{L_0})}~.$$

\noindent
--- on the other hand, the ``position'' of the point $M$, defined by the standard deviation around the mean,
moves from $A$ to $A'$ within the inertial domain (see Fig. \ref{schemabis}) with a typical ``velocity''
$$ \frac{ d x_M}{d \log{r/L_0}} \approx (c_1+1),$$
as $r$ decreases from $\eta_+$ to $\eta_-$.  Within this representation  (see \cite{PhDChev} for details), the
variable $\log{r/L_0}$ may be viewed as ``time''. The correction due to the change of width of $G_{r,L_0}$ is
 neglected. Indeed, this correction expresses as $1/2\sqrt{c_2/\log(r/L_0)}$ and can therefore be omitted
in the near-dissipation range. One then gets
$$ x_{A}-x_{A'} \approx \frac{4}{3} \log (\frac{\eta_+}{\eta_-})  \textrm{ by taking } c_1=\frac{1}{3}\mbox{ .}$$
%\end{itemize}
Eq. (\ref{eq:C2etaplus}) follows immediately. Eq. (\ref{eq:C2etamoins}) can be demonstrated in a similar way by
considering the motion of the point $M$ moving from $B$  to $B'$ within the viscous domain. }

\end{document}